\documentstyle[aps,prl,twocolumn,epsf,graphicx]{revtex}

\begin{document}
\draft
\preprint{}
\twocolumn[\hsize\textwidth\columnwidth\hsize\csname @twocolumnfalse\endcsname
\title{Demonstration Of A Continuously Guided Atomic Interferometer
       Using A Single-Zone Optical Excitation}

\author{M.S. Shahriar$^{1,2}$, Y. Tan$^{2}$, M. Jheeta$^{2}$, J. Morzinksi$^{2}$, 
          P.R. Hemmer$^{2}$, and P. Pradhan$^{1,2}$ }
\address{$^{1}$Department of Electrical and Computer Engineering, Northwestern University, Evanston, IL 60208 \\
      $^{2}$Research Laboratory of Electronics, Massachusetts Institute of Technology, Cambridge, MA 02139} 
\maketitle
\date{\today}
\begin{abstract}
We demonstrate an atomic interferometer in which the atom  passes 
through a \textit{single-zone} optical beam, consisting of a pair of bichromatic 
counter-propagating fields. During the passage, the atomic wave packets in 
two distinct internal states trace out split trajectories, guided by 
the optical beams, with the amplitude and spread of each wave-packet varying 
continuously, producing fringes that can reach a visibility close to unity. 
We show that the rotation sensitivity of this continuous interferometer (CI) 
can be comparable to that of the Borde-Chu Interferometer (BCI). The 
relative simplicity of the CI makes it a potentially better candidate for 
practical applications.
\end{abstract}
]
{\em Introduction} :\,\,\, 
In a typical atomic interferometer[1-8], 
an atomic wavepacket is first split up  by an atomic beamsplitter [9-13], 
then the two  components are redirected towards each other by atomic 
mirrors. Finally, the converging components are made  to interfere
by another atomic beamsplitter. Here, we demonstrate an  
atomic interferometer where the atomic beam is split and 
recombined in a continuous manner. Specifically, in this interferometer, 
the atom  passes through a single-zone optical beam, 
consisting of a pair of bichromatic counter-propagating fields that cause 
optically off-resonant Raman excitations. During the passage, the atomic 
wave packets in two distinct internal states couple to each other 
continuously, and the states each traces out a complicated 
trajectory, guided by the optical beams, with varying amplitudes 
and spreads of each wave-packet. Yet, at the end of the single-zone 
excitation, the interference fringe amplitudes  can reach a visibility 
close to unity. One can consider this experiment as a limiting version 
of the so-called $\pi/2$-$\pi$-$\pi/2$ Raman atomic interferometer, proposed 
originally by Borde [1], and demonstrated by Kasevich and Chu {\em et al.} [2]. 
This configuration is potentially simpler than the Borde-Chu interferometer 
(BCI), eliminating the need for precise alignment of the multiple zones. 
Under  circumstances of potentially practical interest, the CI may
be able to achieve a rotational sensitivity comparable to that of the
BCI, as described later. As such, the relative simplicity of 
the CI may make it an attractive candidate for measuring rotation 
and for other applications. 

{\em BCI and CI} :\,\,\, In order to illustrate the CI, it is 
instructive to  recall the BCI briefly, where the atom is assumed to be a three 
level system in the lambda configuration, with 
two low-lying levels $\vert $a$>$ and $\vert $b$>$, each of which is coupled 
to the level $\vert $e$>$. The atom moves in the \textbf{x} direction 
through two counter propagating laser beams in the \textbf{z} direction, 
which are split in three equidistant zones, as illustrated in Fig. 1 
(top). One of the laser beams couples $\vert $a$>$ to $\vert $e$>$, while 
the other couples $\vert $b$>$ to $\vert $e$>$. The $\pi $/2 Raman pulse in 
the first zone splits the atom into  
\begin{figure}[htbp]
\centerline{\includegraphics[width=8.5cm,height=7.5cm]{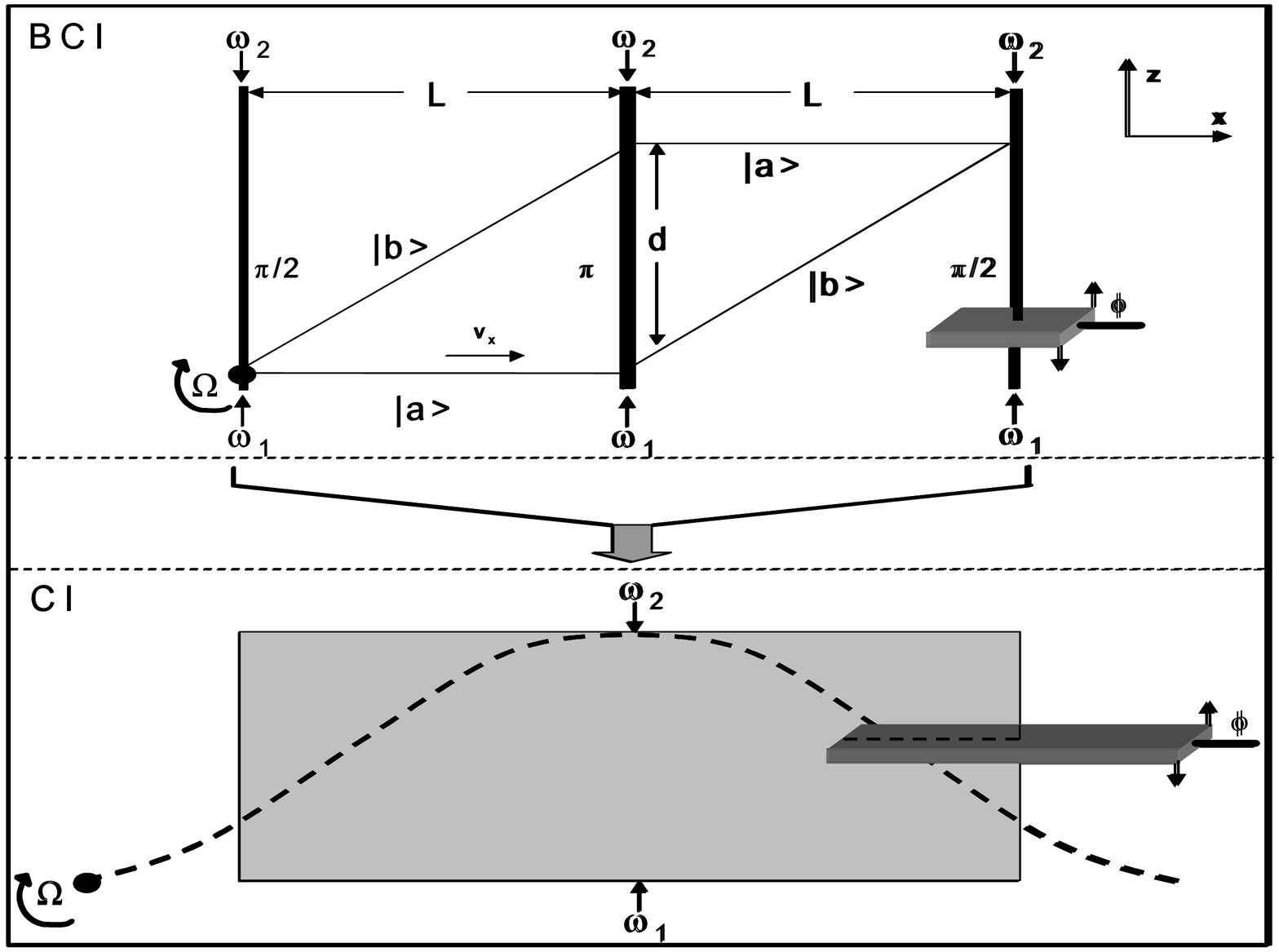}}
\caption{ Schematic illustration of the BCI as compared to the CI. In the BCI, 
the atom passes through three laser pulses of $\pi/2-\pi-\pi/2$ arrangement. 
In the case of CI, the atom passes through one laser pulse of Gaussian shape.
In each case, an external phase $\phi$ is applied by a glass plate that 
rotates along the {\bf x} axis and the interferometer rotates around the 
{\bf z} axis  with an angular velocity $\Omega$.}
\label{fig1}
\end{figure}
\noindent two components: the $\vert $a$>$ part 
travels straight, while the $\vert $b$>$ part picks up a transverse 
momentum of ${2\hbar k}$, where $k$ is the average wavenumber 
of the two laser beams. 

The $\pi $ Raman pulse in the second zone redirects these trajectories, 
followed by a  $\pi $/2 Raman pulse in the last zone, 
which returns the atom to a state where the fraction of the atoms in 
the $\vert $a$>$ state, for example, depends on the phase shift $\phi $ applied 
to this zone, using a glass plate that rotates around the 
\textbf{x} axis. A signal corresponding to the 
population of level $\vert $b$>$ thus varies sinusoidally with $\phi $, with 
the fringe minimum occurring at $\phi $=0 (mod 2$\pi )$. If the whole 
apparatus rotates at a rate $\Omega$ around the \textbf{y} axis, then the 
fringe minimum shifts by an amount given by $\delta \phi =2 \Omega Am/\hbar $, 
where A is the area enclosed by the trajectory of the split components of 
the wave packet and $m$ is the mass of the atom.
 
In contrast, the CI employs only a single zone, for which the profile in the 
${\bf x}$ direction (Gaussian, for example) is chosen to correspond to a 2$\pi$ Raman 
pulse. The phase shift, $\phi $, is applied to a part of this beam, as 
illustrated in Fig. 1 (bottom). For $\phi $=0, it is obvious that the atom 
will be in state $\vert $a$>$ at the end of the interaction, 
just as in the case of the BCI. However, it may not be obvious  how one would 
define an effective area for this interferometer, how the population in state 
$\vert $b$>$, for example, would depend on the phase $\phi $, and what would 
be the rotational sensitivity.

{\em Numerical simulation :}\,\,\, 
We show  theoretically that in fact the functional 
behavior of the CI is very similar to that of the BCI. Here, we report briefly the 
key features of this analysis. A more detailed  analysis is presented in reference [14].
Specifically, we show the calculated trajectories 
of the split components, then determine the  rotational sensitivity, 
and  use it  to determine the effective area of the CI, defined as 
$A_{eff }= \hbar\delta \phi /2\Omega m$.
Finally, we present our experimental results demonstrating the operation of the CI, 
manifested as interference fringes observed as a function of the phase $\phi$. 
The formalism used in this analysis uses a quantized center-of-mass (COM) 
position for the atom along  the $z$ direction. In the electric dipole approximation, 
the Hamiltonian for the system can be written as:
\begin{equation}
\label{eq1}
H=\textbf{P}_z^2 /2m\,+\,H_0 \,+\,q {\bf r}\cdot \left( {{\bf E}_1 +{\bf E}_2 } 
\right),
\end{equation}
where \textbf{E}$_{1}$ and \textbf{E}$_{2}$ are the classical electric field 
vectors of the two counter-propagating lasers, \textbf{P$_z$} is the COM momentum 
in the \textbf{z }direction, $H_0$ is the internal energy, \textbf{r} is 
the position of the electron with respect to the nucleus, and $q$ is the 
electronic charge. We use as the basis  the eigenstates of the non-interacting 
Hamiltonian, $\left| {P_Z } \right\rangle \otimes \left| i \right\rangle 
\equiv \left| {p,i} \right\rangle $.  Define  \textit{$\omega $}$_{i }$ 
as the energy of the $i$th  $(i=a,b,e)$  state, $k_{j}$, \textit{$\omega $}$_{j }$ 
and $ $\textit{$\phi $}$_{j }$ as  the wavenumber, 
the angular frequency and the phase, respectively, of \textbf{E}$_{j}$, 
and $\Omega _{j}$ and $\delta_j$ as the Rabi frequency and the detuning, 
respectively, for the transition excited by \textbf{E}$_{j}$ ($j=1,2)$. 
Considering  counter-propagating laser fields $k_1=-k_2=k$,  
it can be  shown [14] that  this Hamiltonian creates 
transitions only between the following manifold of states for a 
fixed momentum $p$ : $\left| {p,a} \right\rangle \leftrightarrow \left| {p+\hbar k ,e} 
\right\rangle \leftrightarrow \left|{p+2\hbar k ,b} \right\rangle$. 
Let the amplitude of  these three states be $\tilde{\alpha}$, $\tilde{\beta}$, 
and $\tilde{\xi}$ [15-19] within the manifold of a fixed momentum p.
Since the laser beams are far detuned from  resonance, we make the adiabatic 
approximation, which assumes that the intermediate state occupation is 
negligible and that we can set $\dot{\tilde{\xi}}\approx 0$.  
We  then  get the dynamics  of an effective two level system 
$\psi_{eff} \equiv \left[ \begin{array}{c} 
\tilde{\alpha}(p,t)  \\  \tilde{\beta}(p+2\hbar k,t)
\end{array} \right]
\label{c_matrix} $
by solving $i\hbar \dot{\psi}_{eff}=H_{eff}\psi_{eff}$, where  the effective 
Hamiltonian is:
\begin{equation}
\tilde {H}_{eff} \left( p \right)=\hbar \left[ {{\begin{array}{*{20}c}
 {\frac{\Delta }{2}+\frac{\Omega _o }{2}} \hfill & {\frac{\Omega _o }{2}}   
\hfill  \\
 {\frac{\Omega _o }{2}} \hfill & {-\frac{\Delta }{2}+\frac{\Omega _o }{2}}
\hfill  \\
\end{array} }} \right].
\end{equation}
with the effective Rabi frequency $\Omega_0\equiv \Omega_1 \Omega_2/(\delta_1+\delta_2)$ 
and $\Delta\equiv(\delta_1-\delta_2)/2$.  
 
Ignoring any global phase factor which does not depend on p, we can get
the expression for the system wavefuntion at a time $t$ for all $p$:
\begin{eqnarray}
\label{eq4}
\left| {\Psi (t)} \right\rangle =&&\int dp \,\exp(( p^2+(p+2\hbar k)^2 )t /4m\hbar)  \nonumber\\
 && ({\tilde {\alpha } (p,t)\left|{p,a}\right\rangle +\tilde {\beta }(p+2\hbar k,t)
     \left| {p+2\hbar k,b} \right\rangle } )\,. 
\end{eqnarray}
In our analysis of the rotational sensitivity, we  apply this solution 
for the state vector for the case of a laser field with a  Gaussian profile  
in the x direction. The position representation of the wavefunctions for the 
$\vert $a$>$ and $\vert $b$>$ states are then:
\begin{mathletters}
\begin{eqnarray}
\psi _a (x,t)=\int {dp\;\tilde{\alpha}(p,t)\exp (\frac{-ipx}{\hbar })}, \\
\psi _b (x,t)=\int {dp\;\tilde{\beta} (p+2\hbar k,t)\exp (\frac{-ipx}{\hbar })},
\end{eqnarray}
\end{mathletters}
and the probabilities for the atom to be in either state are: 
\begin{mathletters}
\begin{eqnarray}
P\left( a \right)=\int {dp\;\left| {\tilde{\alpha} \left( {p,t} \right)} \right|^2}, \\ 
P\left( b \right)=\int {dp\;\left| {\tilde{\beta} \left( {p,t} \right)} \right|^2}.
\end{eqnarray}
\end{mathletters}
 In order to do a  phase scan in this system, we apply a phase-shift to the 
laser pulse starting from some position $\delta l$ measured from the 
center of the pulse and extending in the direction of propagation of the atom. 
Such a scan can be realized by placing a glass plate in the path of the 
beam, inserted only partially into the transverse profile of the laser beams, 
and rotating it in the vertical direction. 
We  perform our  simulation on a Gaussian field  discretized 
along the x direction, with  the system  rotating with an angular velocity
$\Omega$ during the interaction time $T$. The phase shift for 
this interferometer is  linear for infinitesimal rotations. 
Thus, an effective area for this interferometer can be defined as 
above. We choose to simulate a system with the following parameters: $\Omega 
_{0}$ = 2$\pi $ (7$\times $10$^{4})$ and $l = 3$ $\times $ 10$^{-3}$ m, such 
that $\Omega _{0}$T = 3.3. The atom is a Gaussian wavepacket with a 1/e 
spread of 1/k, where k = 8.0556 $\times $ 10$^{6 }$m$^{-1}$, corresponding 
to the wavelength of the laser, 780 nm. 
The wavepacket centroid trajectories in the CI are shown in Fig 2.[A]  
for $\phi $= 0. Here, the trajectories may appear to be completely 
separated  from one another, however, note that the atomic  
wavepackets are highly overlapped,  given the
width of the wavepacket. The trajectories are plotted in Fig. 2[A] with no 
rotation in the system. If the system is rotating, there will be slight 
deviations in the trajectories, which lead to the rotational fringe shifts. 
Simulations are performed to determine the effective area 
\begin{figure}[htbp]
\centerline{\includegraphics[width=8.8cm,height=7.3cm]{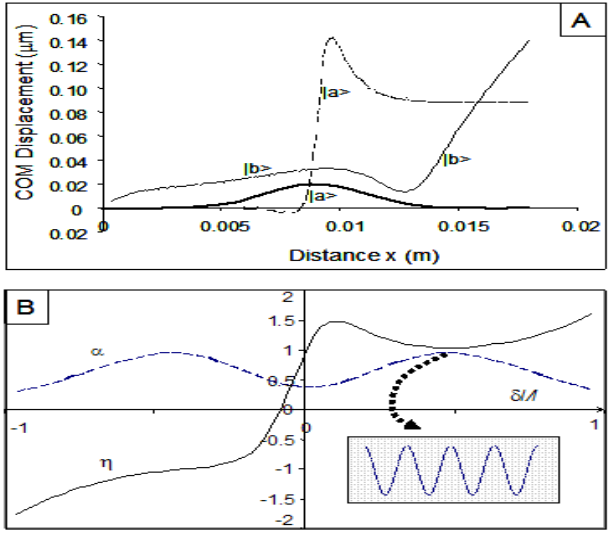}}
\caption{
[A] The trajectory of the split components in the CI, generated from
    numerical simulation. 
[B] The fringe amplitude ($\alpha$) and the normalized effective area 
    ($\eta \equiv A_{eff}/A_0$) vs. $\delta l/l$ for the CI from the same numerical 
    simulation. (Inset) Simulated fringe amplitude vs. 
    external phase $\phi$  at $\eta=.955$.}
\label{fig2}
\end{figure}
\noindent of the interferometer, as a function of the location of the  
the point of application of the phase shift. 

 In order to compare this rotation sensitivity with that of a BCI, we now 
need to know the area of a BCI that would correspond to the parameters of 
our system, the CI. To make this correspondence, note that most of the
interaction in the Gaussian laser profile occurs within one standard 
deviation of the peak of the profile. Thus, it is reasonable to define an 
equivalent BCI with a zone-separation length of L = 3 $\times $ 10$^{-3}$ m 
(so that the three-zone length is 2L), which is the 1/e length of the 
Gaussian profile. The area of a BCI is given by  
$A_0 =L^2 2\hbar k/(mv_x)$. For L = 3 $\times $ 10$^{-3}$ m, we 
get A$_{0}$ = 2.7 $\times $ 10$^{-10}$ m$^{2}$. 

With this value of A$_{0}$ we calculate [14]  the variation of the 
fringe amplitude $\alpha$ and the normalized area ($\eta\equiv A_{eff}/A_0$)  
as a function of  $\delta l/l $.  These  are plotted in Fig. 2 [B]. 
The maximum fringe contrast for our system is 0.955 and occurs at ${\delta \,l} \mathord{\left/ 
{\vphantom {{\delta \,l} l}} \right. \kern-\nulldelimiterspace} l=\pm $ 
0.48. The phase scan displaying this result is shown in the inset. The quality 
factor is approximately one for $\vert {\delta \,l} \mathord{\left/ 
{\vphantom {{\delta \,l} l}} \right. \kern-\nulldelimiterspace} l\vert $ 
$>$ 0.25, which means that if the phase is applied starting in this range of 
values, \textit{our CI interferometer will provide the same 
rotation sensitivity as a BCI of the same size does} [14].

{\em CI Experiment :}\,\,\, 
Our experimental setup is shown schematically in Fig. 3[A]. A thermal $^{85}$Rb 
atomic beam is collimated by two apertures, each of radius 50 $\mu $m and 
seperated by 112 mm. The interaction region is magnetically shielded by 
$\mu$ metal, with a magnetic bias field along the direction of the Raman beams, 
provided by  Helmholtz coils. 

The Ti-sapphire laser used in the experiment is locked to the F=3 
$\rightarrow$ F$'$=3 transition using a saturated absorp- 
\begin{figure}[htbp]
\centerline{\includegraphics[width=8.5cm,height=7.3cm]{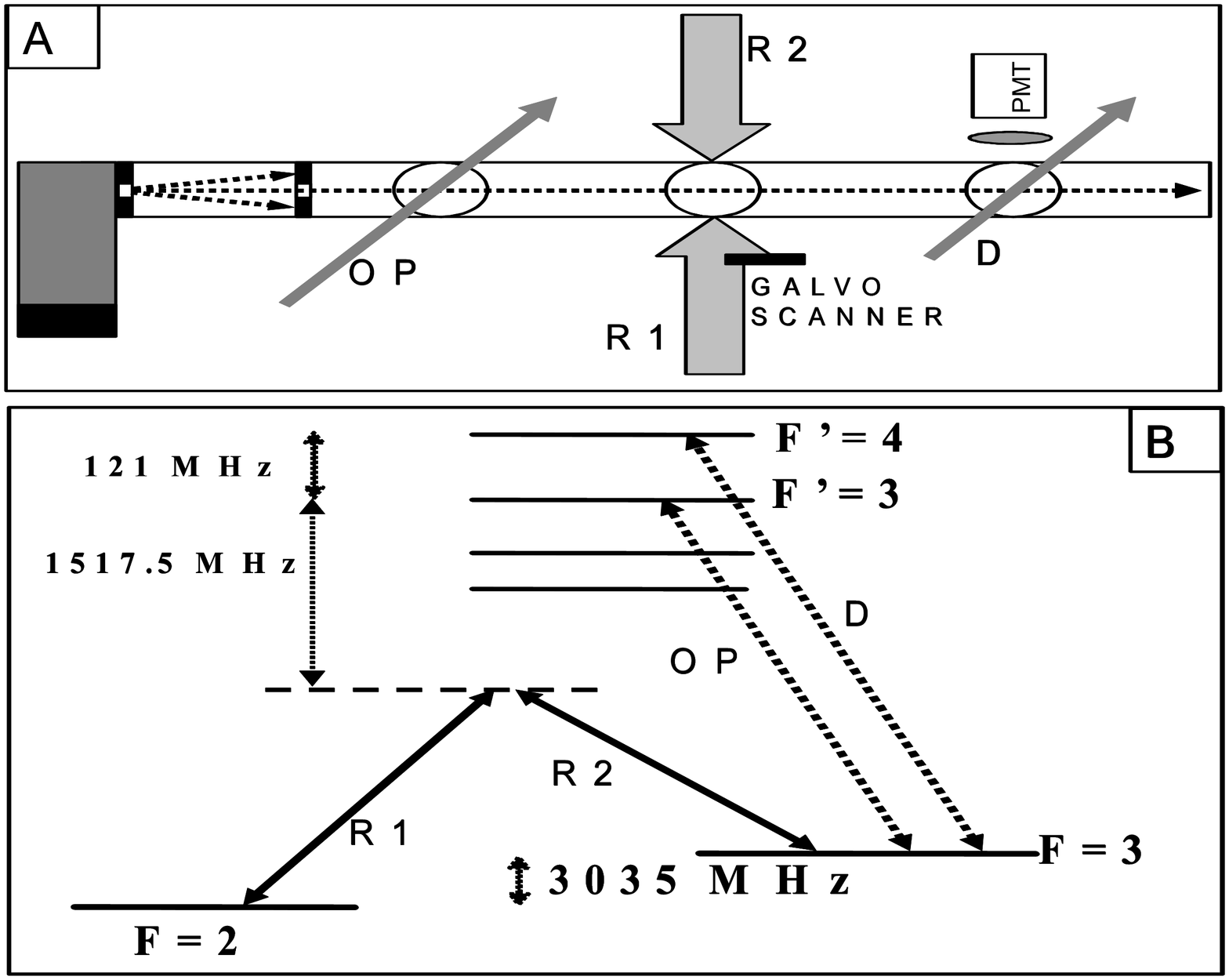}}
\caption{ 
[A] Schematic illustration of the experimental setup of our  CI.
Collimated thermal  $^{85}$Rb atoms  are first  optically pumped to  prepare 
an initial state. Next, the atoms pass through an interaction 
zone containing a bias magnetic field and  two Raman pulses ($R_1$ and $R_2$). 
A glass plate on a galvo scanner is inserted in the edge of  $R_1$. 
[B] Schematic illustration of the $^{85}Rb$  transitions employed in 
realizing the CI. Raman transition is  realized between the ground states of 
5$^2$S$_{1/2}$(F=3) and 5$^2$S$_{1/2}$(F=2) by two  Raman beams $R_1$ and $R_2$. 
To prepare an initial state, the F=3 ground state is optically pumped 
(OP) to the F=2 ground state via the state 5$^2$P$_{3/2}$(F$'$=3). 
The population of the F=3 state is detected (D) via a cycling transition to 
the 5$^2$P$_{3/2}$(F$'$=4) state, by a PMT. }
\label{fig3}
\end{figure}
\noindent tion cell. A beam at this frequency ( OP in Fig.3[B] ) is used 
to prepare the atom in the F=2 sate via optical pumping. The detection beam
(D in Fig. 3[B]) is generated by using an AOM at 120 MHz.
To generate the Raman beams, another part of the  laser is 
divided into two parts with a 50/50 beam splitter. 
One part (R$_1$) is upshifted through a 1.5 GHz AOM, and the other part 
(R$_2$) is downshifted through another 1.5 GHz AOM.  The two 1.5 GHz AOMs are controled 
by the same microwave generator, and so are phase-correlated  with each other.
The downshifted beam is red-detuned by 1.5 GHz from the F=3 to F$'$=3 transition,
and the upshifted beam  is red-detuned by 1.5 GHz from the F=2 to F$'$=3 transition. 
See Fig. 3[B] for a level diagram of the frequencies. Detection is performed by 
collecting the fluorescence using a photomultiplier tube (PMT). 
The bias field separates the magnetic sublevels (F=2 and F=3 have opposite 
sign g-factors) so that the Raman transition between F=2, m$_F$=0 and
F=3, m$_F$=0 is shifted from the neighboring Raman transitions
(e.g., F=2, m$_F$=1 and F=3, m$_F$=1) by nearly 1 MHz/Gauss.  Given the 
transverse velocity spread which corresponds to a Raman linewidth of about 
3.2 MHz FWHM (including transit time broadening), the use of a bias field of 
2 Gauss enables us to resolve the sublevel Raman transitions. The experiment 
is performed by exciting the m$_F$=0 to m$_F$=0 transition.

 To scan the phase of the interference of the CI, we insert 
a 1mm thick glass plate  mounted on a galvo scanner into an edge of a Raman beam. 
The galvo is  mounted on a magnetic base and is driven by a BK Precision 
function generator. The beam that passes though the glass 
plate aquires a phase that varies with  the angle of the plate. 
The intensity of the detected signal  varies as the phase 
is scanned, and a plot of the observed population in state F=3 is shown in Figure 4[A]. 
To calibrate the phase scan produced by the glass plate, we perform  a 
Mach-Zehnder optical interferometer experiment with the same   scanner.  
Figure 4[B] shows the Mach-Zehnder interference pattern  varying  
with the scanned phase. 
The data shown in Fig. 4 [A] is averaged over 512 traces, with a galvo scan 
rate of 32 Hz. The amplitude of the background (not shown) is  about 5 times that 
of the fringe amplitude. This is  mostly due to a combination of two factors:
imperfection of the optical pumping in the first zone and residual optical
pumping during the Raman interaction into F=3 state. No attempt is  made to 
optimize the signal to noise ratio for this proof of principle experiment.
\begin{figure}[htbp]
\centerline{\includegraphics[width=9cm,height=6cm]{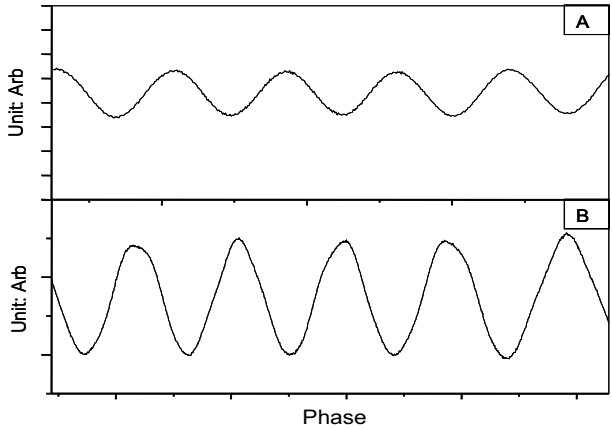}}
\caption{[A] The observed atomic interference fringes in the CI,
 produced when a galvo scanner rotates a glass plate which produces a phase shift.
[B] The corresponding optical fringes in  a Mach-Zhender interferometer using the same  scanner.
Here, one full fringe corresponds to an optical path length of $\lambda$(=780nm).}
\label{fig4}
\end{figure}
{\em  Discussion :}\,\,\,
The CI may be  operationally simpler because it uses only one zone. In 
practice, this means that there is no need to ensure the precise parallelism of the 
three zones, as needed for the BCI. Therefore, the CI may be preferable to the 
BCI, given that the effective area and therefore the rotational sensitivity 
of the CI can be close to that of the BCI. One potential concern is that 
while the BCI can accommodate an effective length (separation between the 
first and the third zones: 2L) as long as several meters, such a long 
interaction length for the CI would be impractical. On the other hand, an 
interferometer that is several meters long is unsuited for practical usage 
such as inertial navigation. Therefore, it is likely that a practical 
version of the BCI would be much shorter (several cm's) in length and 
would reduce the rotational sensitivity drastically. 
This concern can potentially be overcome by using a slowed 
atomic beam (e.g., from a magneto-optic trap or Bose-condensate), so that 
the transverse spread of the split beams would be much larger, thereby 
compensating for the reduction in the longitudinal propagation distance. 
Under such a scenario, the CI would be simpler than the BCI, while yielding 
the same degree of rotational sensitivity.

This work was supported by DARPA grant No. F30602-01-2-0546 under the QUIST 
program, ARO grant No. DAAD19-001-0177 under the MURI program, and NRO 
grant No. NRO-000-00-C-0158.


\vspace{-.4cm}
\begin{thebibliography}{50}
\bibitem{} C.J. Borde, Phys. Lett. A \textbf{140 }(10), 1989.
\bibitem{} M. Kasevich and S. Chu, Phys. Rev. Lett. \textbf{67}, 181 (1991).
\bibitem{} L. Gustavson, P. Bouyer, and M. A. Kasevich, Phys. Rev. Lett. \textbf{78}, 2046 (1997).
\bibitem{} M. J. Snadden, {\em et al.}, Phys. Rev. Lett. \textbf{81}, 971 (1998).
\bibitem{} T.J. M. McGuirk, M. J. Snadden, and M. A. Kasevich, Phys. Rev. Lett. \textbf{85}, 4498 (2000).
\bibitem{} Y. Tan, J.  Morzinski, A.V. Turukhin, P.S. Bhatia, and M.S. Shahriar, Opt. Comm. {\bf 206}, 141 (2002).
\bibitem{} D. Keith, C.Ekstrom, Q. Turchette, and D.E. Pritchard, Phys. Rev. Lett. \textbf{66}, 2693 (1991). 
\bibitem{} D.S. Weiss, B.C.  Young, and S. Chu, Phys. Rev. Lett. \textbf{70}, 2706 (1993).
\bibitem{} T. Pfau {\em et al.}, Phys. Rev. Lett. \textbf{71}, 3427 (1993).
\bibitem{} U. Janicke and M. Wilkens, Phys. Rev. A  \textbf{50}, 3265 (1994). 
\bibitem{} R. Grimm, J. Soding, and Yu.B. Ovchinnikov, Opt. Lett. \textbf{19}, 658 (1994).
\bibitem{}  T. Pfau, C.S. Adams, and J. Mlynek, Europhys. Lett. \textbf{21}, 439 (1993). 
\bibitem{} K. Johnson, A. Chu, T.W. Lynn, K. Berggren, M.S.  Shahriar, and M.G. Prentiss, Opt. Lett. {\bf 20}, 1310 (1995).
\bibitem{} M.S. Shahriar, M. Jheeta, Y. Tan, P. Pradhan, and A. Gangat, quant-ph/0309147. 
\bibitem{} P.M. Radmore and and P.L. Knight, J. Phys. B. \textbf{15}, 3405 (1982).
\bibitem{} M. Prentiss, N. Bigelow, M.S. Shahriar and P. Hemmer, \textit{Optics Letters}, \textbf{16}, 1695(1991). 
\bibitem{} P.R. Hemmer, M.S. Shahriar, M. Prentiss, D. Katz, K. Berggren, J. Mervis, and N. Bigelow, 
           Phys. Rev. Lett. \textbf{68}, 3148 (1992). 
\bibitem{} M. S. Shahriar and P. Hemmer, \textit{Phys. Rev. Lett.}, \textbf{65}, 1865(1990). 
\bibitem{} M. S. Shahriar, P. Hemmer, D.P. Katz, A. Lee and M. Prentiss, \textit{Phys. Rev. A}, 
           \textbf{55}, 2272 (1997).
\end{thebibliography}
\end{document}